\documentclass[twoside]{dis08}
\usepackage[latin1]{inputenc}
\usepackage[dvips]{graphicx,epsfig,color}
\usepackage{wrapfig,rotating}
\usepackage{amssymb,amsmath,array}
\def\GeV{{\rm GeV}}

\begin{document}
\title{Parton Distributions and QCD at LHCb}

\author{R.S. Thorne$^1$\footnote{Royal Society University Research Fellow}, 
A.D. Martin$^2$, W.J. Stirling$^2$ and G. Watt$^1$ 
%
%
\vspace{.3cm}\\
%
1- Department of Physics and Astronomy, University College London,
WC1E 6BT, UK
%
\vspace{.1cm}\\
2- Institute for Particle Physics Phenomenology,
University of Durham, DH1 3LE, UK
}

\maketitle

\begin{abstract}
We consider the impact that can be made on our understanding of parton 
distributions (PDFs) and QCD from early measurements at the LHCb experiment. 
The high rapidity values make the experiment uniquely 
suited to a detailed study of small-$x$ parton distributions and hence will 
make a significant contribution towards the clarification of 
both experimental and theoretical uncertainties on PDFs and their
applications.
\end{abstract}

\section{Introduction}

\begin{wrapfigure}{r}{0.48\columnwidth}
\vspace{-1.6cm}
\centerline{\includegraphics[width=0.48\columnwidth]{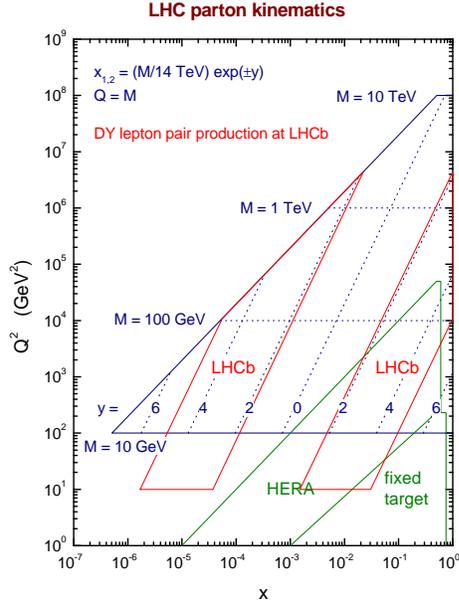}}
\vspace{-0.8cm}
\caption{The kinematic range of $x$ and $Q^2$ which is probed at the 
LHC.}\label{Fig:LHCbkin}
\end{wrapfigure}

The kinematic range for particle production at the LHC is
shown in Figure \ref{Fig:LHCbkin}. The $x$ values of the PDFs are 
$x_{1,2}=x_0 \exp(\pm y),$ where $x_0 =M/\sqrt{s}$.
Smallish $x \sim 0.001-0.01$ parton distributions are therefore 
probed by the standard production 
processes at the LHC, e.g. $W,Z$, at central rapidity. However, LHCb, which 
has a rapidity coverage of $1.8< y < 4.9$, automatically probes the PDFs at 
very low (and high) values of $x$. At the low values the experiment 
constraints begin to run out and there are also potentially 
significant theoretical uncertainties.

\begin{figure}
\vspace{-1.4cm}
\centerline{\includegraphics[width=0.5\columnwidth]{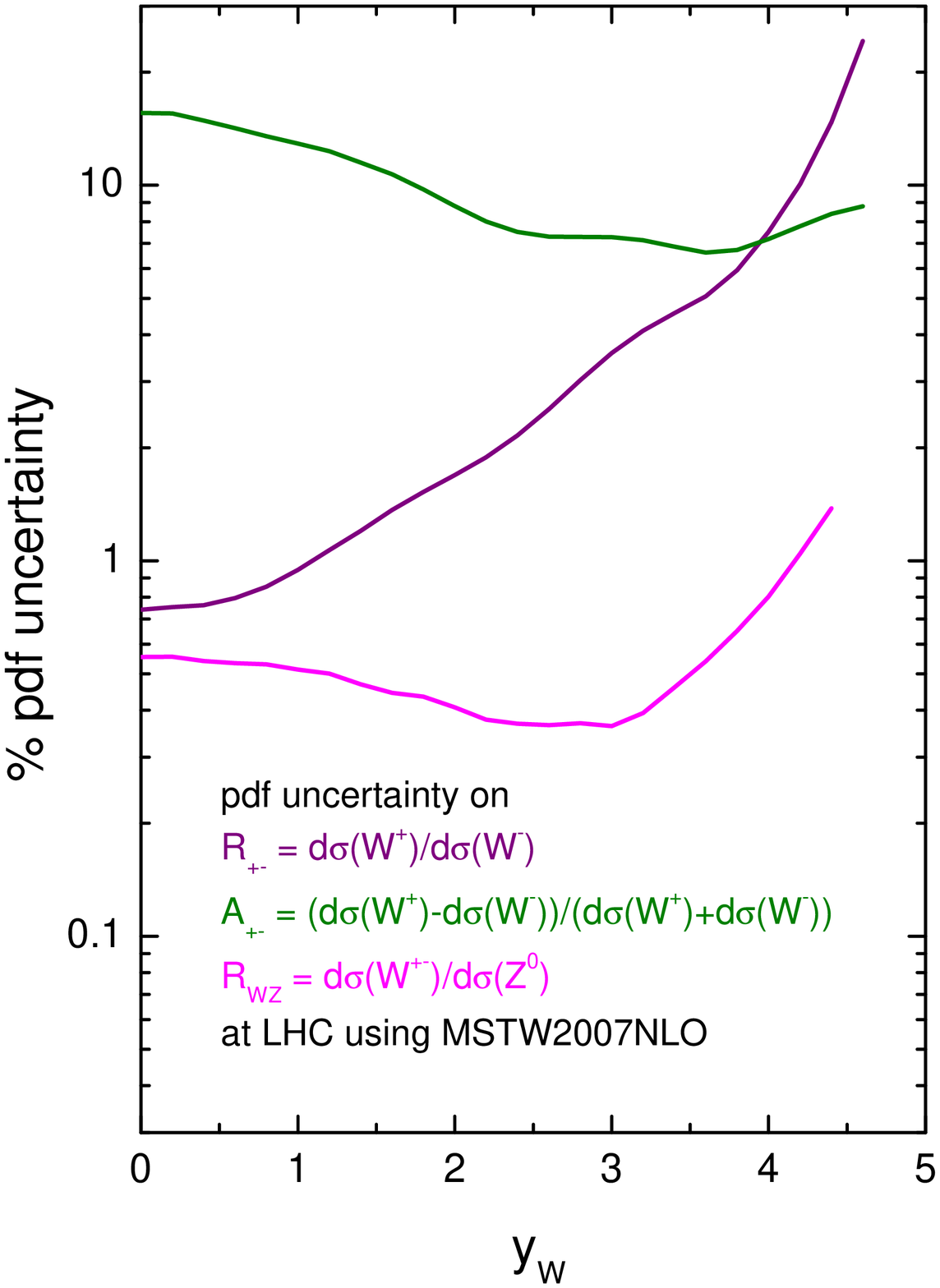}
\includegraphics[width=0.5\columnwidth]{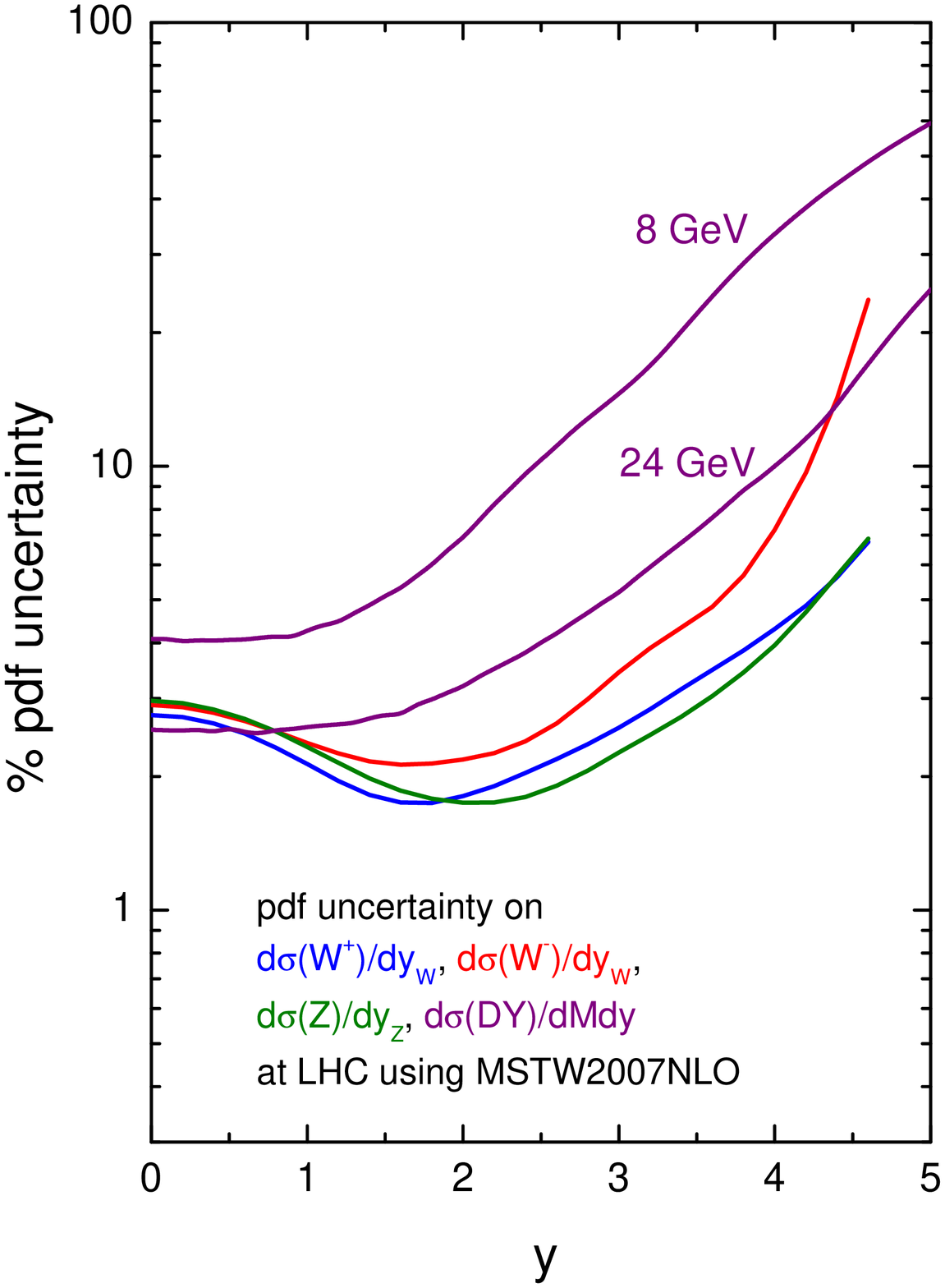}}
\vspace{-1.0cm}
\caption{The uncertainties on cross-section ratios and total cross-sections 
at the LHC.}\label{Fig:crossunc}
\vspace{-0.4cm}
\end{figure}

We can gain useful information on PDFs and QCD from both total cross-section 
measurements and ratios of cross-sections at LHCb \cite{slides}. 
The $90\%$ confidence level limit uncertainies using the preliminary 
MSTW 2007 PDFs \cite{dis2007}  are shown in Figure \ref{Fig:crossunc}. 
The results are exact but 
they can be understood more easily using some simplifications. 
Since $x_1 > x_2$, particularly at high values of $y$, and since 
sea quarks die away at high $x$, we assume 
$q_1(x_1)\bar q_2(x_2) + \bar q_1(x_1)q_2(x_2) 
\approx q_1(x_1)\bar q_2(x_2).$ Also at small $x_2$ we assume that 
$\bar u(x_2) = \bar d(x_2)$. We finally assume that the 
cross-sections are dominated 
by up and down contributions, which is reasonable for obtaining 
general results. Using these simplifications we obtain for the 
cross-section ratios the expressions
\begin{equation}
R_{Z/W} \simeq \frac{A_u \, u(\tilde x_1)+
A_d \, d(\tilde x_1)}{u(x_1)+d(x_1)}, \quad A_{\pm}  \simeq \frac{u_V(x_1)-
d_V(x_1)}{u(x_1)+d(x_1)}, \quad R_{\pm} \simeq \frac{d(x_1)}{u(x_1)}, \nonumber
\end{equation}
where $A_u = v_u^2+a_u^2$ and $A_d = v_d^2+a_d^2$, i.e. the sum of the 
squares of the vector and axial-vector couplings. The use of $\tilde x_1$ 
illustrates that $x$ values are slightly higher for $Z$ production than 
for $W$ production.
In particular the large uncertainty on $R_{\pm}$ for high $y$ 
reflects that on the 
high-$x$ down quark distribution. 
Similarly for total cross-sections we have 
\begin{eqnarray}
\sigma_{W^{+}} &\simeq &    u(x_1)\bar d(x_2), \quad 
\sigma_{W^{-}} \simeq     d(x_1)\bar u(x_2), \\ \nonumber
\sigma_{Z} \, & \simeq &   
A_u u(x_1)\bar u(x_2)+A_d 
d(x_1)\bar d(x_2),\quad  
\sigma_{\gamma^{\star}} \simeq 
4/9 \,u(x_1)\bar u(x_2)+1/9\, d(x_1)\bar d(x_2). \nonumber
\end{eqnarray}
The uncertainty on $\sigma_Z$ and $\sigma_{W^+}$ at high $y$ is dominated by 
the valence up quark and the sea quarks for $x \sim 0.0001$, both 
of which are limited to a few percent, hence the small uncertainty in 
Figure \ref{Fig:crossunc}. The uncertainty on $\sigma_{W^-}$ is dominated by 
the down quark for large $x_1$, as for $A_{\pm}$. 
For the virtual photon, 
$x_{1,2} = (M_{\gamma}/14,000\,\GeV)\exp{(\pm y)}$, so if 
$M_{\gamma} \leq 20\,\GeV$, $x_2$ can be $\leq 0.00001$ 
and $x_1$ stays away from the valence uncertainty at high $x$.
The increased uncertainty when these extremely small $x$ values are probed 
is evident in the figure. 

\section{Potential results from LHCb}

\begin{wrapfigure}{r}{0.38\columnwidth}
\vspace{-0.2cm}
\centerline{\includegraphics[width=0.36\columnwidth]{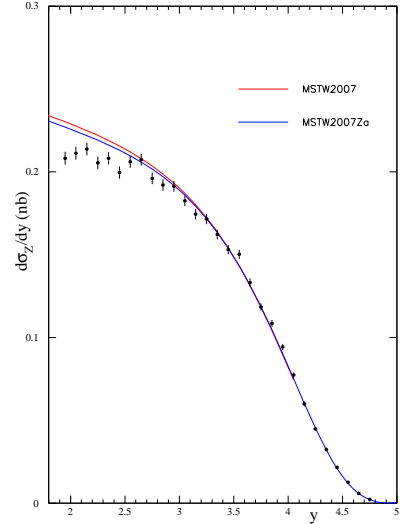}}
\vspace{-0.4cm}
\caption{The result of fitting to hypothetical 
$d\sigma_Z/dy$ data.}\label{Fig:ztot}
\end{wrapfigure}

Depending on the cross-section under investigation, early data can either 
reduce the uncertainty on PDFs or test the validity of the order-by-order 
perturbative expansion. 
With $1 {\rm fb}^{-1}$ of data LHCb \cite{ronan} expects to obtain 
data for $Z \to \mu^+\mu^-$ for $1.8 < y < 4.9$ with $\sim 1 \%$ precision
(ignoring luminosity uncertainty) if presented in bins of width $0.1$. 
This is sensitive to the small-$x$ quarks and gluon distribution for 
$x < 0.001$, but the most precise data will be in the range where 
HERA data \cite{HERA}
already provide very good constraints. However, if the data are a different 
shape to the prediction they can highlight problems with fixed order
calculations. Illustrated in Figure \ref{Fig:ztot} is pseudo-data shifted 
by a mulitplicative factor $0.05(y-3.4)$ compared 
to the current NLO prediction. The lower (blue) line shows the result of a 
new fit. It is not possible to obtain good agreement, i.e. 
$\chi^2 =103/30$. Hence,
this type of result could 
imply that small-$x$ resummations are important, or that 
higher twists/saturation is contaminating the current extraction of small-$x$ 
PDFs from  HERA data. 

\begin{figure}[h]
\vspace{-0.4cm}
\centerline{\includegraphics[width=0.4\columnwidth]{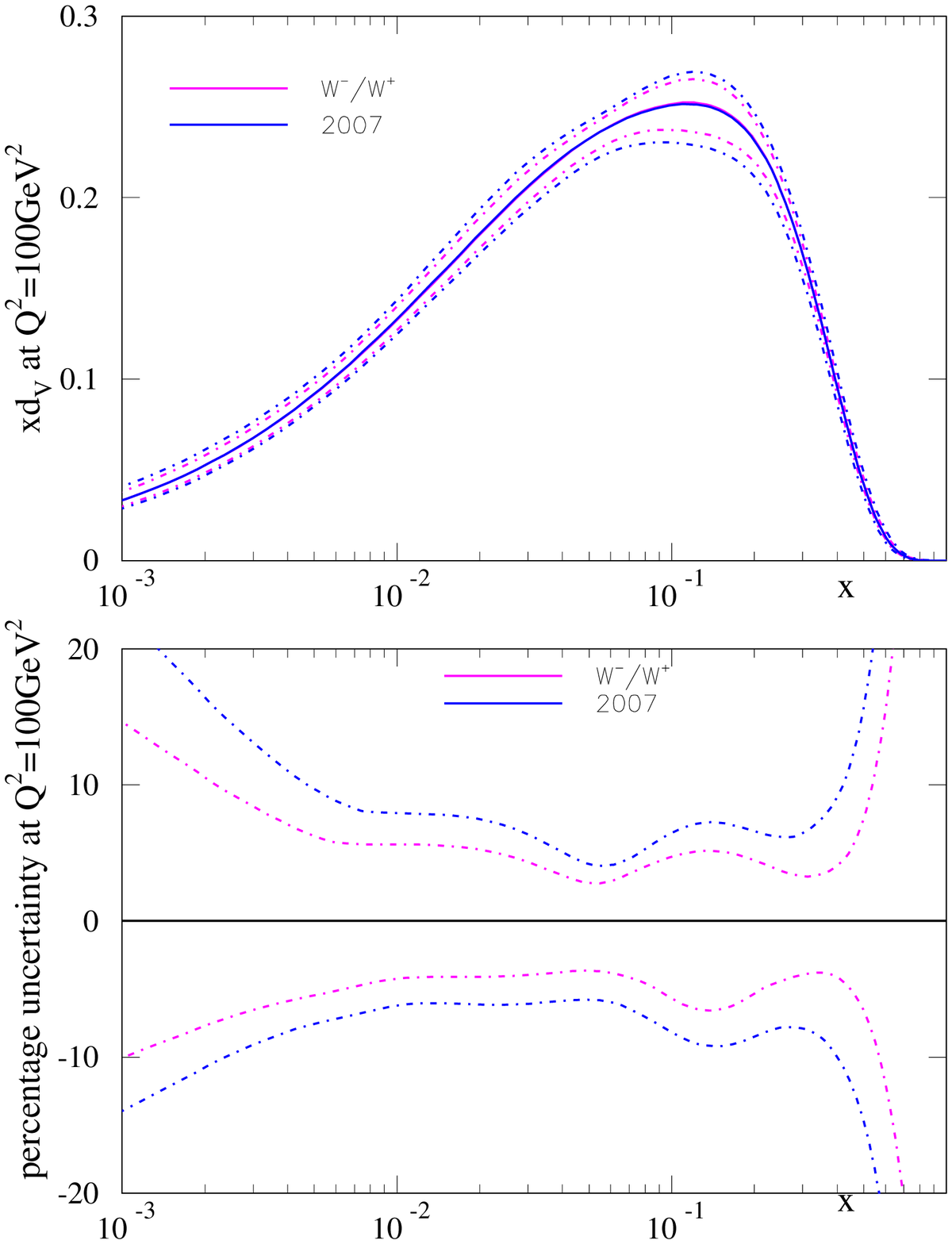}
\hspace{1cm}
\includegraphics[width=0.4\columnwidth]{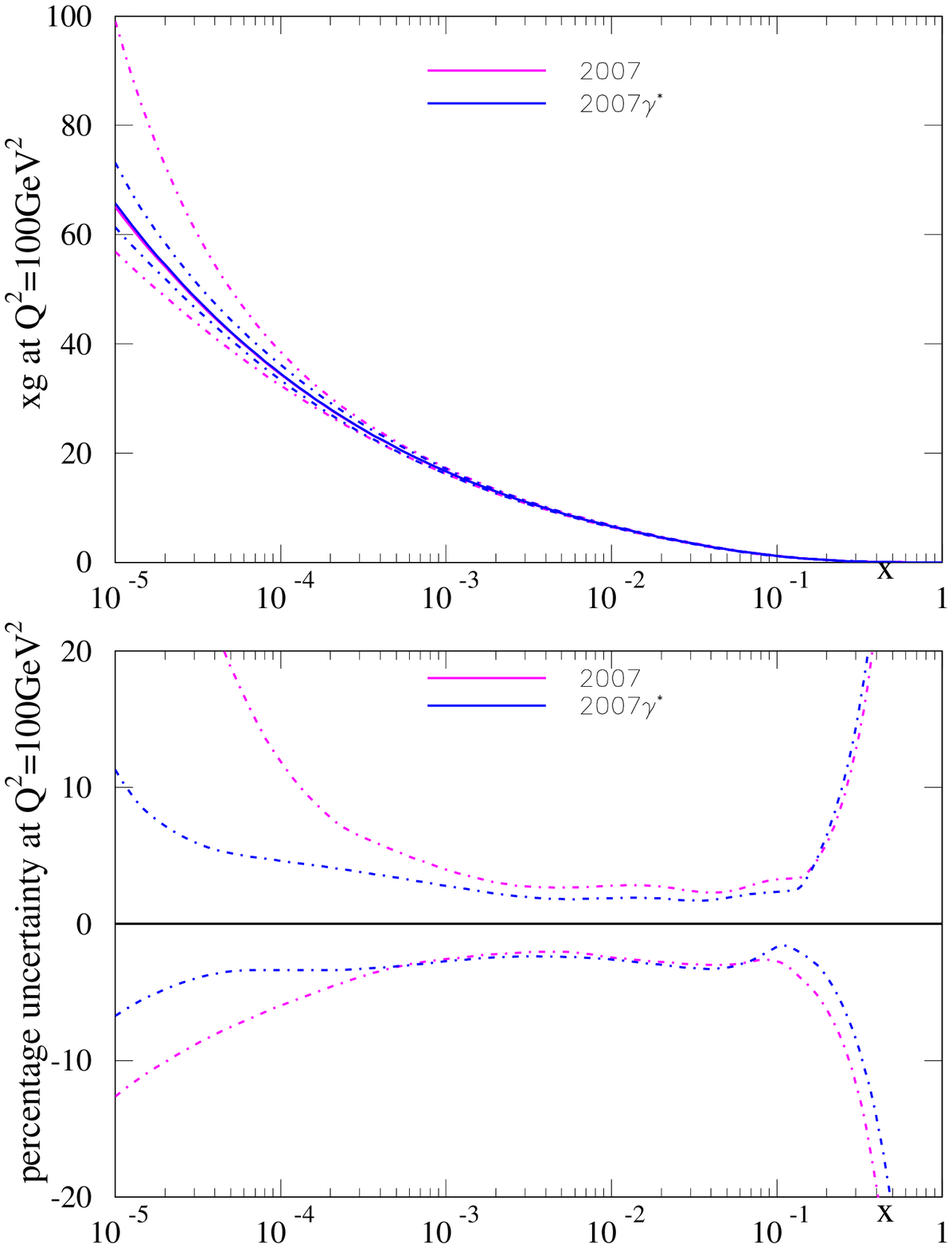}}
\vspace{-0.2cm}
\caption{The improved constraint on $d_V(x,Q^2)$ from fitting 
to $(d\sigma_{W^-}/dy)/(d\sigma_{W^+}/dy)$ data (left) and on $g(x,Q^2)$ 
from fitting to $(d\sigma_{\gamma^{\star}}/dy)$ for $M_{\gamma^{\star}} = 
14\,\GeV$ (right).}\label{Fig:partunc}
\end{figure}

Consider instead the example of 
$1 {\rm fb}^{-1}$ data on $\sigma(W^-)/\sigma(W^+)$. (In practice 
this will be measured as lepton asymmetry, but very similar 
considerations apply.) 
The cross-section for $W \to \mu\nu_{\mu}$ is ten times that for
$Z \to \mu^+\mu^-$, but is more difficult to measure with 
more systematics. We have 
assumed similar uncertainties, which is perhaps conservative \cite{ronan}. 
In this case, 
assuming  the data are consistent with the prediction they immediately become 
one of the main constraints on the down valence distribution, as shown in 
Figure \ref{Fig:partunc}. Again a discrepancy between data and theory could 
easily highlight shortcomings, but it is not very obvious what these might be 
in this case (except 
perhaps an unexpected large $\bar u, \bar d$ asymmetry at small $x$.)

Even more dramatic is the constraint that could potentially be made by a 
measurement of the low-mass Drell-Yan cross-section. In principle the 
cross-section for e.g. a $14\,\GeV$ virtual photon is similar to that  for
the on-shell $Z$ but the background is far more of a problem. We assume 
(perhaps slightly optimistically) that 
for $1 {\rm fb}^{-1}$ 
the uncertainty at the lowest $y$ is about twice that for 
$Z \to \mu^+\mu^-$ but the cross-section (and hence statistics) 
falls off less quickly with $y$ because the valence quark fall off is not 
reached. Data of this type for a single
mass value produce the extra constraint shown in the right of 
Figure \ref{Fig:partunc}. Since at these values of $x$ the quarks are driven 
by evolution from the gluon, the constraint is very similar for small-$x$ 
quarks. 

\begin{wrapfigure}{r}{0.5\columnwidth}
\vspace{0.7cm}
\centerline{\includegraphics[width=0.45\columnwidth]{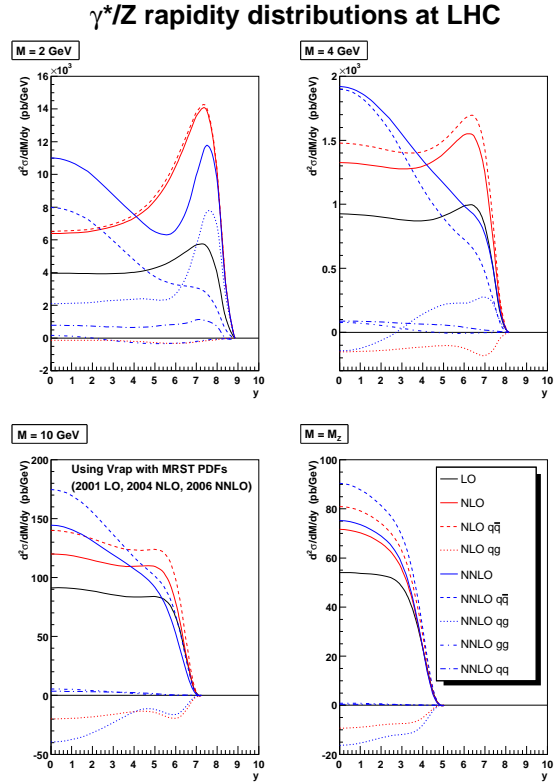}}
\vspace{0.7cm}
\caption{The Drell-Yan cross-section as a function of rapidity in
various mass bins.}\label{Fig:dyplot}
\end{wrapfigure}

Since for $x < 0.0001$ the uncertainty on PDFs is already large, a 
measurement of the low-mass virtual photon cross-section in 
one mass bin is more inclined to give an improved constraint on PDFs 
rather than test
the theory, unless it is very different from the prediction (more so than 
required for Z production). However, 
measurements in the range of mass bins will be an excellent test of 
small-$x$ evolution. We already have reason to believe that the perturbative 
series is unstable, where use of the available NNLO matrix elements 
\cite{FEWZ} shows that for $M_{\gamma^{\star}} < 20 \GeV$ there is sizable 
variation in predictions as one goes from LO to NLO to NNLO, as shown in 
Figure \ref{Fig:dyplot}. The instability can be traced to the increasing 
divergence of terms in the perturbative expansion as one increases the 
order of the calculation. This has a significant effect on both the 
small-$x$ PDFs and the matrix elements for the cross-section \cite{DY}. 
Measurements
on low-mass Drell-Yan production at LHCb, and also at ATLAS and CMS, will be 
important for our understanding of the convergence of the perturbative 
series in QCD. Alternative measurements, e.g. heavy flavour production, could
produce complementary information, but are less clean. 
 
To summarise, suitable measurements at the LHC experiment will place extra 
constraints on PDFs and QCD which will then feed back into predictions for 
all quantities. LHCb has a unique reach, and precision measurements at high 
rapidity can probe some of the most 
important and interesting regions of parton kinematics, with even 
limited data from early running making a major impact.


\begin{footnotesize}

\end{footnotesize}
\end{document}